# The Missing Ones: Key Ingredients Towards Effective Ambient Assisted Living Systems

Hong Sun, Vincenzo De Florio, Ning Gui and Chris Blondia
*PATS Research Group, University of Antwerp, 1 Middelheimlaan, Antwerp, 2020, Belgium.*
*Interdisciplinary Institute for Braodband Technology (IBBT),Gaston Crommenlaan 8, Gent, 9050, Belgium*
*Email: {hong.sun, vincenzo.deflorio, ning.gui, chris.blondia}@ua.ac.be*

**Abstract.** The population of elderly people keeps increasing rapidly, which becomes a predominant aspect of our societies. As such, solutions both efficacious and cost-effective need to be sought. Ambient Assisted Living (AAL) is a new approach which promises to address the needs from elderly people. In this paper, we claim that human participation is a key ingredient towards effective AAL systems, which not only saves social resources, but also has positive relapses on the psychological health of the elderly people. Challenges in increasing the human participation in ambient assisted living are discussed in this paper and solutions to meet those challenges are also proposed. We use our proposed mutual assistance community, which is built with service oriented approach, as an example to demonstrate how to integrate human tasks in AAL systems. Our preliminary simulation results are presented, which support the effectiveness of human participation.

## 1. Introduction

As well known, the proportion of elderly people keeps increasing since the end of last century. The European overview report of Ambient Assisted Living (AAL) investigated this trend [21] aiming to find out an efficient solution to help the elderly people independently living.

AAL aims at extending the time the elderly people can live in their home environment by increasing their autonomy and assisting them in carrying out activities of daily lives by the use of intelligent products and the provision of remote services including care services. Most efforts towards building ambient assisted living systems for the elderly people are based on developing pervasive devices and use Ambient Intelligence [4] to integrate these devices together to construct a safety environment. Living assistance systems and assistive devices are thus developed to facilitate the daily lives of these elderly people. Such efforts are close to achieve the goal of assisting the elderly people to live independently by transferring the dependence from human side to assistive devices. However, what we deem as their biggest limitation is that although they are to achieve such goal, in so doing they also reduce the social connections of the assisted people and hinder human participation. Although recent researches have now noticed the importance of human participation and intend to increase social connections, there is still no effective way to enhance the human participation in ambient assisted living systems.

In this paper, first we review the existing efforts of bringing human participation into ambient assisted living systems. Then, we observe how the common approaches draw a dividing line between those who are to be cared of and those who take care. By doing so, AAL services based on this axiom confine the users to a passive role that discriminate and exclude them from the better part of society, which introduces frustration and discomfort. We deem avoiding or minimizing such frustration and discomfort from the user side as the biggest challenge towards the success of ambient assisted living systems. Our solution to meet such challenge is by changing the roles that people played in ambient assisted living systems. Instead of classifying people into assisted people and caregivers, we propose to construct the ambient assisted living system as a mutual assistance community where every subscriber is equally treated as peer

participant – each with their own diverse abilities and capabilities, but all contributing to the common welfare. We expect such change of roles could help to bring in more human participation and social interaction in ambient assisted living. We also expect such change could help to save the social resources while also providing services to the elderly people in an effective way.

Other challenges of bringing human participation, especially coping with the dynamical natural of human services are also discussed in this paper, and we conjecture that the dynamic characteristics of the service orientated approach provide a satisfactory solution to meet such challenges. To substantiate our claim and provide a practical example, we introduce the mutual assistance community, an architecture where services from human side and services from machine side are integrated together as web services under the management of a service coordination center, where a web service matching engine is implemented.

An example of organizing group activity where people join as peer participants is given in this paper. A simple simulation model of a mutual assistance community is also presented: preliminary results suggest the effectiveness and efficiency of bringing human participation in ambient assisted living.

## 2. Bringing human participation in AAL systems

Much research were carried out on building intelligent environments around people, which aimed at providing assistive services in pervasive environments, constructing a better environment and providing people with better lives [7] [13]. Recent research efforts also aim at bringing intelligibility in ambient environments to model the human behaviours [9], thus providing better services to the end users. Services provided by those projects are promising to help the elderly people to ease their lives and keep them safe by monitoring some of their health status. However, services provided in those projects are similar and lack of diversity. What is more important, the scenarios in these projects are still not complete enough to meet elderly people's needs in their daily lives and help them maintain satisfactory independent lives.

The Amigo project [6], though not specifically designed to assist the elderly people, investigated ambient intelligence for the networked home environment to provide attractive user services and im-prove end-user usability. Pervasive devices are managed in the Amigo project in an adaptive, context-aware and autonomous way. The scenarios of this project proved that systems such as this are able to provide their users with customized services. The applications are not restricted to the home environment, but extended to include the work environment through mobile devices, and are also able to connect family members together.

As such, the Amigo project is a huge step towards general introduction of the networked home and towards Ambient Intelligence as it enhances considerably the usability of a networked home system and expands the application domain. The achievements made in the Amigo project could be applied to Ambient Assisted Living for the elderly people to provide services based on advanced ICT technology. However, as the Amigo project is not specifically designed to assist the elderly people living independently, still there are unmet challenges to fully express the potential of adopting those technologies to assist the elderly people living independently and safely in their houses. What is even more important, Amigo services lack of human participation and the communication between the assisted people and the community is predefined and restricted to their family or neighbours. This inherently restricts services discovery inside a limited group and may isolate the user from the "world outside."

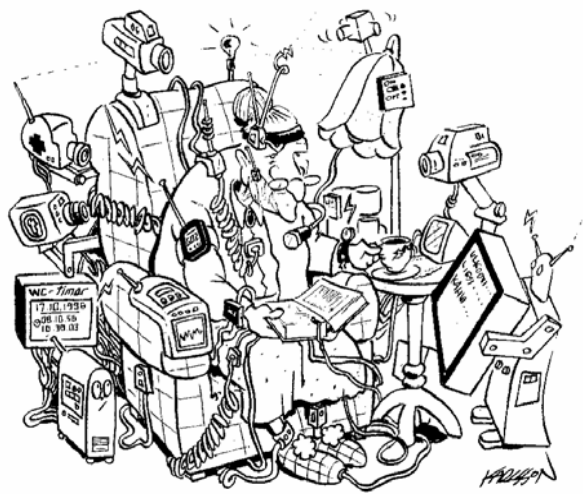

Fig. 1. Side-effect of over-using assistive devices [15]

It is our conviction that human participation is essential in building effective ambient assisted living systems. The AAL country report of Finland remarked that "the (assistive) devices are not useful if not combined with services and formal or informal support and help" [5]. We share this view and deem informal caregivers from relatives, friends and neighbouring people as indispensable when constructing timely and cost-effectively services to assist the elderly people. The usage of assistive devices helps to transfer the dependence from human side to machinery side, thus establishing some degree of independence. However, the dependence on the assistive devices unconsciously reduces the social connections of the assisted people. We feel that, without communication with the outside world, elderly people assisted by those assistive devices are only safely surviving rather than actively living. Figure 1 shows the possible side-effect of over-using technology without proper human participation. Although the effect is exaggerated, the picture reminds us that we should be cautious not to leave the elderly people only with assistive devices, but also with our sympathy, compassion, communication, and help.

Luckily, many researchers are now realizing the importance to include social intelligence and social connection [3]. They have made efforts toward building social links between the elderly people and their families or neighbours to address their needs and increase their social connections. However, those solutions are still mainly focused on the technical aspect and human participation is also very limited. We argue that effective and efficient solutions to meet the AAL challenges should combine the forces from both the technological part and the societal ones:

  o  On one hand, the involvement of human beings could help fully express the potential of smart devices, and maintain the social awareness of the elderly people.
  o  On the other hand, the usage of advanced ICT technology could better connect the elderly people together, e.g. organizing community activities.

In the following sections, we firstly discuss the challenges of developing pervasive home-care services exploiting human participation [25], and then we propose a possible approach to construct such a system.

## 3. Challenges of bringing human participation in AAL systems

The previous sections stated the promising approaches to help the elderly people living safely and independently in their own houses. We also proposed to integrate the human services into AAL systems to enrich the available services and create a less intrusive environment. However, there are still many challenges towards the implementation of such an environment. In this section, we will discuss those we deem to be the main challenges, the necessary steps, and some possible solutions to effectively deliver services which exploit human participation in AAL systems.

### 3.1. Dynamicity of service availability

Although informal caregivers may help to reduce the needed social resources, and increase the social connections, they are also very difficult to be utilized, due to their inherently dynamic nature: the availabilities of these services are continuously changing. How to manage this dynamicity becomes a big challenge.

We believe Service Oriented Architecture (SOA) could be a good approach to cope with the above mentioned dynamicity. SOA is a flexible, standardized architecture that supports the connection of various services, and as such it appears to be an ideal tool to tackle the dynamicity problem. The application of SOA, such as the OSGi platform [18], can also help to establish a framework such that various smart devices could be integrated together and could be automatically discovered, called, started and stopped. The unified service format will also help the processing of service matching. Research efforts of using OSGi to build safety home environment are also reported in [4].

Another attractive feature of SOA can be found in recent researches towards integrating people activities into service frameworks, which culminated in two specifications launched in the summer of 2007: Web Services Human Task (WS-HumanTask) [27] and WS-BPEL Extension for People (BPEL4People) [8]. WS-HumanTask targets on the integration of human beings in service oriented applications. It provides a notation, state diagram and API for human tasks, as well as a coordination protocol that allows interaction with human tasks in a service-oriented fashion and at the same time controls tasks' autonomy. A people activity could be described as human tasks in the WS-HumanTask specification. The BPEL4People specification supports a broad range of scenarios that involve people within business processes, using human tasks defined in the WS-HumanTask specification. These two specifications could help to meet the challenges of integrating human services in the SOA framework of the proposed mutual assistance community.

### 3.2. Service mapping

How to let the computer automatically map the available/requested services is a big challenge towards constructing effective AAL systems.

The foundation for service mapping is service description. A Semantic Knowledge Base is required to precisely describe the advertised services: certain ontology libraries describing the domain knowledge of the home-care environment should be developed with the interdisciplinary collaboration of the researchers in this domain. With such domain knowledge, conceptual models for semantic service matching could be applied. OWL-S [19] is currently the most used technology in this domain; it is able to provide a framework for semantically describing web services from several perspectives, e.g., service inquiry, invocation, and composition.

There are several service matching tools developed for matching OWL-S services, such as OWL-S Matcher [26], OWL-S UDDI/Matchmaker [20], and OWLS-MX Matchmaker [17]. The main drawback of the first two is that the matching process takes a large amount of time, while the biggest limitation of the latter one is its being memory intensive [11]. In our opinion, these tools serve as good starting points to investigate AAL web service matching, while we believe more elegant and efficient matching engines should be developed. We have made some preliminary tests of service matching in home-care service matching – details can be found in Section 4.2.

### 3.3. People's willingness to participate

People's willingness to participate in AAL systems needs to be investigated and encouraged – how to encourage people joining e.g. a mutual assistance community (see Section 4) is a big challenge.

In [2] [3], the persuasiveness of ambient intelligence is discussed: in order to achieve human participation, AAL system should provide both trust persuasion and enough motivation to get people's acceptance. As we pointed out in the previous section, it is our conviction that any effective AAL system cannot leave aside the contributions coming from society itself, in all forms, with the active participation of informal caregivers, professionals, and even the elderly people themselves. In order to encourage more people to make contribution to AAL system, we need to understand their drives to provide help to others, and stay active in the community. The main drive for people to help others is not merely money, but also includes moral duty and their social image. One main reason that keeps people active in an online community is to build up a good image for their avatars and win respect from other community residents. An AAL system with participation of informal caregivers could also reward the informal caregiver in this way. Social studies to stimulate people to work as volunteers should be thoroughly carried out.

Besides encouraging people to help the ones in need as informal caregivers, how to encourage the elderly people to join such a community is also a challenge. We would ascribe the elderly people's unwillingness to use assisting system from two folds – psychological one and technological one. In the rest of this section we introduce these two aspects in details and explain how we made efforts to reduce this unwillingness.

### 3.3.1. Psychological frustration:

When people are getting old, a relevant source of frustration comes from losing physical strength, but what torches them most lies probably in the psychological sphere: they are becoming passive consumers of the societal services rather than active producers. In so doing, they also lose their self-esteem. Almost all the AAL systems for the elderly people consider their users as people who are weak and require to be passively assisted by others. For the designers of such systems, being able to maintain some degree of independence without bringing too much burden to our society appears as an already ambitious goal. However, those systems neglect the fact that the elderly people can still make their contributions to our society through their valuable experiences. A home-care system with human participation could help to encourage the elderly people to actively participate in group activities as peer participants, and possibly even to use their experiences to help the younger generations to solve, e.g., some of their work and school problems [25]. We expect these activities could help the elderly people to find themselves still useful and thus enable them to live in more active ways. Such possibilities will be discussed in more details in the following section.

### 3.3.2. Technological frustration:

Elderly people are usually scared by the application of new technology. In order to help them get used to the ambient assistive devices, we should construct user friendly interfaces, and also provide appropriate trainings to their users. Developing adaptive, natural and multimodal human computer interfaces is the main challenge of future interfaces in assisted living [16]. It is also suggested to get people involved in how to use the assistive devices before they really need them [10].

In the solution proposed in next section, elderly people do not only benefit from keeping connected with other people, but also are provided chances to make contributions to society, to feel that they are still useful, and to live in an active way with self-esteem. Such benefits could provide our elderly people with stimulations to break the technical barrier.

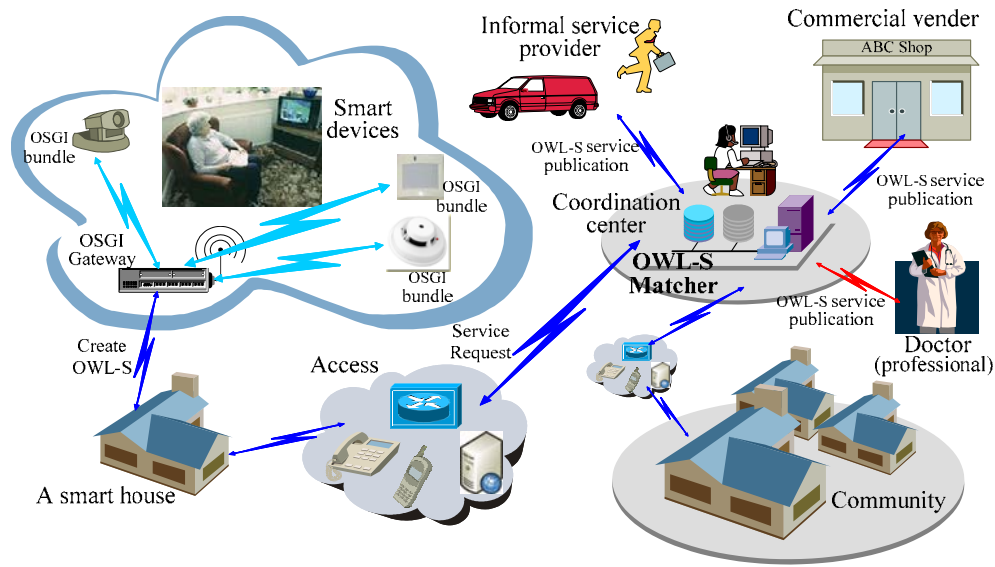

Fig. 2. Organization of mutual assistance community

## 4. Mutual assistance community – a user-centered approach

### 4.1. Community organization

The previous section discussed the challenges of developing AAL systems with human participation. In this section, we propose to construct a so-called "mutual assistance community" to bring services from human side into AAL environments. Such a system integrates the services from human beings with the applications provided by assistive devices, and best utilizes the available resources providing services to the people in need in an effective way [12] [22] [23]. Figure 2 shows the organization of our mutual assistance community. Assistive devices are deployed to construct a smart house environment managed by a local coordinator to build up a safety environment around the assisted people.

As mentioned already, the most important asset integrated in our community is indeed the people themselves. Our proposed community allows disparate technologies and people to work together to help people who suffer from aging or disabilities. People who are able to provide services are encouraged to do so and assist the requesting people as informal caregivers. Elderly people are also encouraged to participate in the group activities, which not only helps to maintain physical and psychological health but also reduces the requests of professional medical resources. Professional caregivers (such as doctors, specialists etc.) are included in the community to provide emergency and professional medical service. Commercial vendors are also included, which brings convenience to the user and diversifies the service type, at the same time laying the foundation for economical exploitation and self-sustainability.

In fact, the importance of the informal caregiver is also notified by previous researches: for instance, Aware Home [7] points out that technology should support networks of formal and informal caregivers, and the scenario of Amigo project [6] also shows informal caregivers could help to provide first- aid help to neighbouring people. However, the link between the informal caregivers and the elderly people are statically fixed there. Our proposed community can flexibly connect the needed help and available informal caregiver services through web service publication, matching and binding. The elderly people can also use this approach to initialize and join group activities, and inter-generational activities could also be carried out in this way. During the inter-generational activities, the younger generation could help the elderly people on physical demanding tasks as informal caregivers. Though physically weak, the elderly people accumulated valuable experiences and knowledge during their lives. They may use such knowledge to assist the younger generation. During this process, not only the younger generation gets their needed answer, the elder generation also finds an access to make their contribution to our society. The elderly people may find themselves still useful, stand with more active living attitude, thus avoiding the frustration of feeling useless.

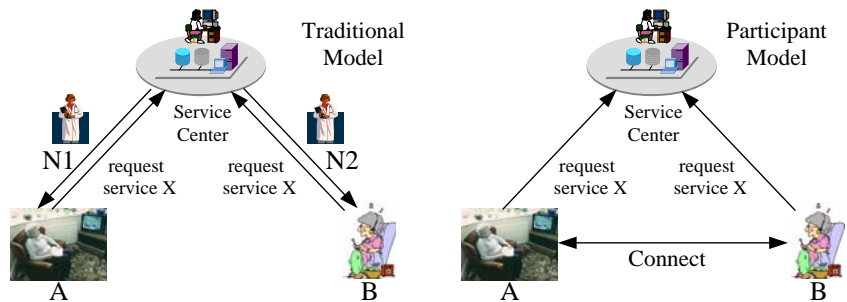

Fig. 3. Organize group activities with "Participant Model"

In our mutual assistance community, we also introduced a so-called "Participant Model" to organize group activities. The concept is that elderly people may be interested in and should be encouraged to take some group activities, e.g. walking in the park together, playing chess etc. Traditionally, such requests are satisfied by either informal caregivers or nursing people. But in our mutual assistance community, the community will try to connect the elderly people to take group activities together (as shown in Fig. 3). In this way, people are not passively receiving help, but actively joining those activities as peer participants, which not only saves the social resources, but also contributes to their psychological health.

### 4.2. Service processing structure

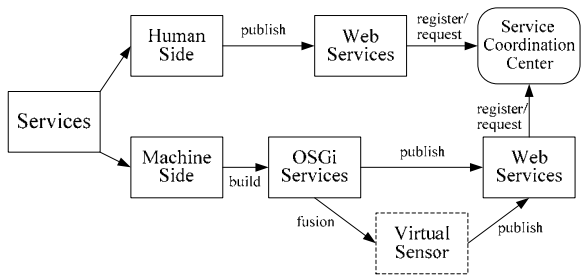

Fig. 4. Structure of service organization

The service structure of our mutual assistance community is shown in Fig. 4. Services from assistive devices and human tasks are processed separate-ly and integrated together. Services from assistive devices are built as OSGi services (bundles), and coordinated by the OSGi gateway. Part of these OSGi services will be published directly as web services in the gateway, while the others will be firstly merged together into composite indicators to abstract higher level information and then published as web services. We call such services "virtual sensors."

On the human side, service processing is straighter forward. Human services are published as OWL-S web services; the availability of the informal caregivers is registered in the service coordination center while the request from the elderly people is also posted as service requests. A user could be registered as a service consumer and a service provider at the same time, based on the type of services. For instance, an elderly person could request physical challenging services, while providing knowledge or experience-based advisory services.

Through the above approach, the available services and the requested services, no matter if coming from the human or machine side, are all represented as web services in the service coordination center. In our implementation, the process of service match is executed by OWL-S Matcher, which is able to reflect different matching types, e.g. totally match, subsume (one instance is the sub-class of another instance), etc.

Fig. 5 shows a fragment of the advertised service in OWL. The #Entertainment ontology (service type) and the #Informal Provider ontology (service provider) are declared in the my_ontology.owl file and specified as inputs and outputs of the service. This is because the OWL-S Matcher could assign different matching acceptance for inputs and outputs, and separating the service provider and service type enables the service requester to specify different acceptable matching degrees for service type and service provider separately.

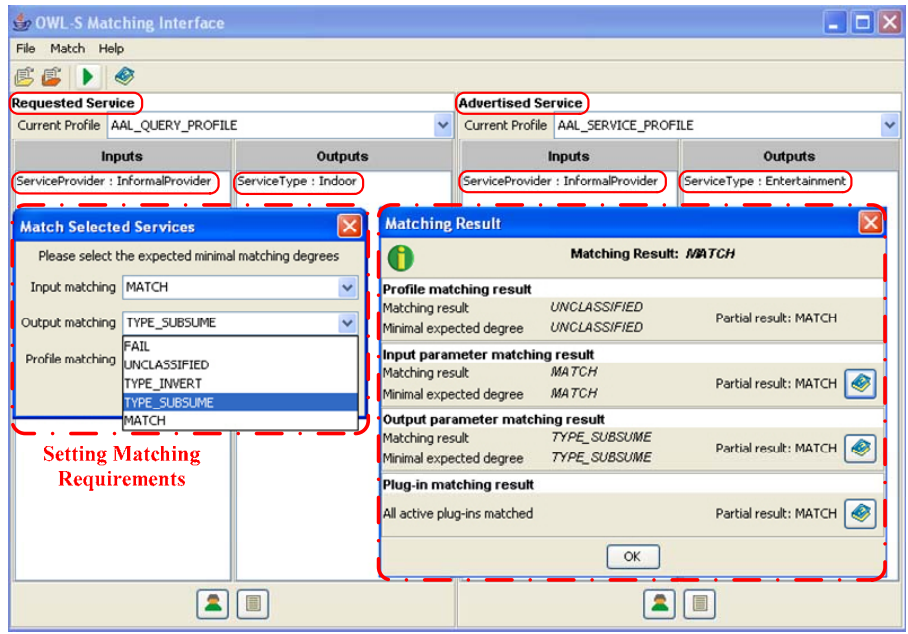

Fig. 6. OWL-S Matcher interface and matching result

```
</profile:textDescription>
<profile:hasInput  rdf:resource="#_InformalProvider"/>
<profile:hasOutput rdf:resource="#_Entertainment"/>

<profile:has_subclass rdf:resource="AAL_SERVICE_PROCESS" />

<profile:has_process rdf:resource="AAL_SERVICE_PROCESS" />
</profile:Profile>

<process:ProcessModel rdf:ID="_AAL_SERVICE_PROCESS_MODEL">
<service:describes rdf:resource="#AAL_SERVICE_SERVICE"/>
<process:hasProcess rdf:resource="#AAL_SERVICE_PROCESS"/>
</process:ProcessModel>

<process:AtomicProcess rdf:ID="AAL_SERVICE_PROCESS">
<process:hasInput  rdf:resource="#_InformalProvider"/>
<process:hasOutput rdf:resource="#_Entertainment"/>
</process:AtomicProcess>

<process:ServiceType rdf:ID="_Entertainment">
<process:parameterType rdf:resource=
"http://127.0.0.1/ontology/my ontology.owl#Entertainment" />
    <rdfs:label></rdfs:label>
</process:ServiceType>

<process:ServiceProvider rdf:ID="_InformalProvider">
<process:parameterType rdf:resource=
"http://127.0.0.1/ontology/my ontology.owl#InformalProvider"/>
    <rdfs:label></rdfs:label>
</process:ServiceProvider>
```

Fig. 5. Fragment of the advertised service in OWL

Figure 6. shows the interface of the OWL-S Matcher. In the window on the left side, user may specify different matching degrees (exact match, subsume, etc). In the window on the right side, the matching result is displayed. The example shown in Fig. 6 shows the matching process between an available *entertainment service* provided by *informal caregiver* and a request for *indoor service* provided by *informal caregiver*. The service types and providers

described in web service may only use the ontology specified in the ontology library. The matching of the service provider yields an exact match, while the matching on the service type comes as subsume, as the requested one (*indoor service*) is defined beforehand as sub class of the *entertainment service* in the ontology library. However, as *subsume match* in service type also meets the requirement of the user – the user expected at least a *subsume match* on the service type – the final result for this matching process is *match*.

## 5. Scenario and simulations

To better illustrate the functions of our proposed mutual assistance community, we will firstly use a scenario to show how the community organizes group activities. Afterwards, we will use a simple model to mimic the overall performance of a mutual assistance community, and present our preliminary simulation results.

### 5.1. Scenario: participation in group activities

Mary is 70 years old and lives alone in Antwerp, Belgium. In the afternoon of a sunny day, Mary would like to have a walk with someone in the Middelheim Park, which is close to her home. Mary de-cides to use the mutual assistance community to find someone who also wants to have a walk in the park. She switches on the TV, which is the graphical interface of the mutual assistance community. She navigates the service menu, which is built as an ontology tree, and selects the "Group Activity". A few photos will be presented to her, representing group activities such as chatting, jogging, walking together, etc. Mary chooses the icon corresponding to walking; she types in the location she wants to hold this event as "Middelheim Park". For the service organization, Mary chooses to receive service from peer "participants"; she specifies the deadline as "today, 8pm".

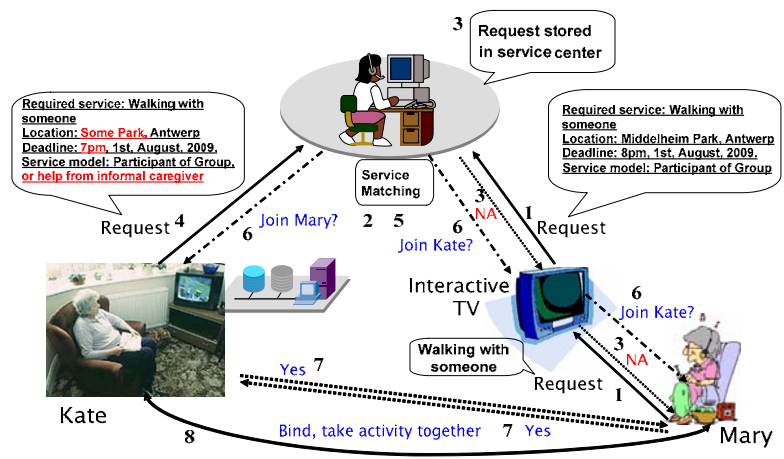

Fig. 7. Scenario of "Participant Activity"

After making such inputs, the service matching engine, which located in the service coordination center, starts searching for the appropriate services. The settings for the search, translated from Mary's actions, are as follows:

Required Service: Walk with someone.
Location: Middelheim Park, Antwerp.
Deadline: 8pm, 1st, August, 2009.
Service Form: Participant of Group Activities.

Unfortunately, there is no matched service for Mary's request; the system sends this result to Mary, and keeps her request in the database before the deadline expires. A few hours later, Kate submits a similar request. The system checks the request from Kate, and finds that it matches Mary's request to initiate a walking activity. Thus an information dialog pops onto Mary's screen: "*Kate* initiates a *walking* activity in *somewhere* by *7pm today*; if you agree to take this *group activity* with *Kate*, press the **OK** button and we will forward your contact information to *Kate*."

Mary agrees to join this activity, she presses "OK" on her remote controller, while at the other side, Kate also accepts the participation of Mary. Their contact information is then displayed on their TV screens respectively, they call each other to confirm the time and place to meet, and later have a nice time walking together in the park.

In this scenario, Mary and Kate take the group activity, walking in park, together. No additional help is required to meet their requests, so that social resources are effectively saved, and they also avoid the possible frustration of having to be taken care by others. Figure 7 illustrates the actions taken in this scenario; the displayed numbers indicate the steps of the indicated action, the indicator "NA" refers to service not available, and the other lines in red indicate that the system may find more matches by exploiting the hierarchical relationships of the compared objects. We are currently working on the graphic user interface to implement the above mentioned scenario.

## 5.2. Simulations

In order to evaluate the effectiveness of mutual assistance community, we have set up a simple cell grid model to imitate the community behavior. The characteristic of a mutual assistance community is that services from informal caregivers are highly dynamic, i.e. the availability of services is continuously changing. Our simulation model reflects such dynamicity by constructing an $n \times n$ cell grid; each cell in this grid represents an individual, and the whole grid represents a community. Figure 8 shows the graphic interface of the simulation model. The simulation model simplifies the categories of the requested ser-

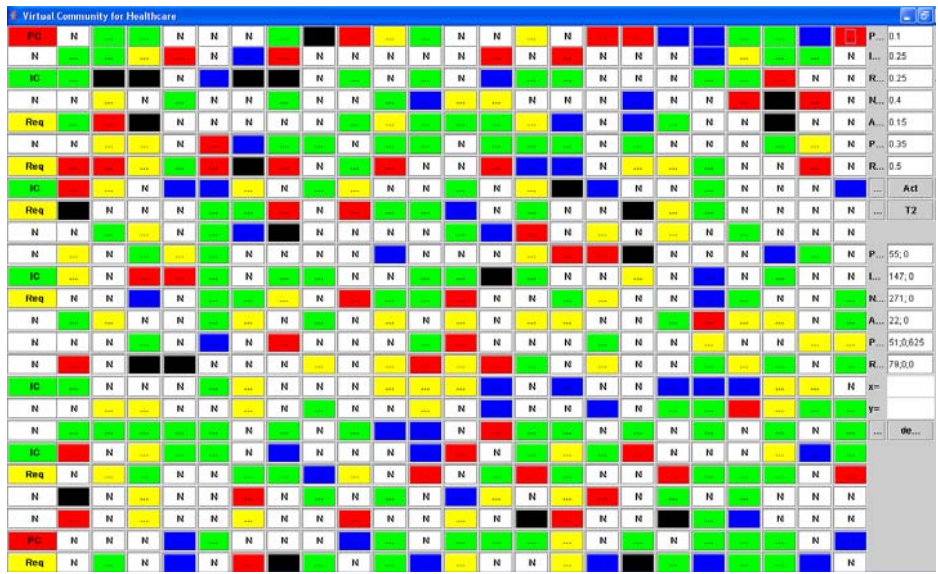

Fig. 8. Simulation model of mutual assistance community

vices and available services, dividing the community dwellers into the following categories:

a) Professional Caregiver: those who are able to provide professional medical care.
b) Informal Caregiver: those who are willing to provide help on non-medical services.
c) Neutral: those who neither need service, nor provide it.
d) Service Requester: those who need services, they are further divided into three sub categories:
   1) Alarm: requests need to be served by professional caregiver.
   2) Normal Request: requests could be served by either informal or professional caregiver.
   3) Participant: requests to join group activities as peer participant.

The above mentioned community dwellers are represented as cells with different colours in our simulation model in Fig. 8. The community shown in Fig. 8 changes in discrete time steps, and each activity (e.g. receives/provides service) takes a certain number of steps to finish. The needed steps to complete an activity depend on the workload of the requested service, which is a randomly generated number ranging from 5 to 25. On every step, cells who are receiving services will decrease their workload value by 1. Once their workload value becomes 0, it indicates the activity is finished, those who have received their services will be set as neutral cell, and those who have provided their services become available again. Every step, those cells that are inactive (cells not connected with other cells on service activities) will change their role with a certain probability. For instance, an informal caregiver may change into neutral, which mimics the event that an informal caregiver becomes not available.

In order to simplify the simulation model, a cell in the community may only look up the states of its neighbouring cell, and once it may establish a connection with its neighbour to provide or receive service, such a connection will be established, and it will last certain steps for the service delivery. In the upper right side of the simulation model in Fig. 8, the rate of different cells, and the probabilities when a cell changes its state can be predefined. The simulation results are accumulated, calculated and displayed in the lower right part of Fig. 8.

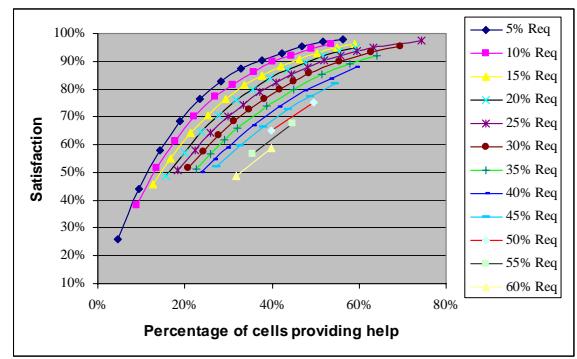

Fig. 9. Performance with informal caregiver

The simulations take into account the averaged successful rate to receive a service in time and the steps of service delay that a user may experience. Our results demonstrate that the presence of informal caregivers helps reduce the social resources and provide daily assistance timely. Figure 9 shows how many client cells can receive timely help through the introduction of informal caregivers. The multicoloured curves represent the different rates of the client cells in the initialization state, from 5% in the top to 60% in the bottom. The degree of satisfaction of the client cells indicates the percentage of the client cells that received help from their neighbours timely after their requests. A provider can serve only one client cell in this simulation. Fig. 9 shows that providing a certain number of informal caregivers, many of the requests could be provided immediately [22].

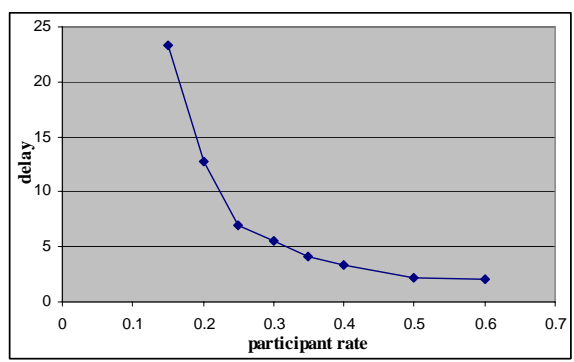

Fig. 10. Performance with "Participant Model"

When the elderly people are actively participating group activities, the dependence on social resources are further reduced, both the failures and service latencies (i.e., the time difference between service requests and their response, if any) goes down considerably. Fig. 10 shows that when the participant rate is 15%, the average latency is 16.3 time steps; when the participant rate increases to 25%, the averaged latency drops to 7.0; and when the participant rate reaches 60%, the averaged latency drops to 2.0. The failures of service delivery also go down with the increase of participants; its trend follows the service delay in Fig. 10. More details and results of our simulations could be found in our previous paper [23].

The simulations also demonstrate that satisfactions of the people who are requesting services are largely depending on the dwellers' willingness to provide help or participating to group activities. Another feature is that the size of the community does not bring direct effect on the users' satisfaction, however expanding the contact that a user may reach with the other users may help to increase users' satisfaction rate. The results also suggest that a stable community with less changes on people's roles may create higher satisfaction.

## 6. Conclusions

This paper discussed some crucial issues and prerequisites for building effective AAL systems for elderly people. We have observed the research efforts of building pervasive home-care environments with advanced Ambient Intelligence, which promises to provide safe environments around the elderly people in their own houses. However, we also foresee some challenges of such solutions: in particular we consider a concrete threat – possible social isolation – due to the over-use of technology and lack of communication between the assisted people and the outside community. We believe effective and efficient solutions to assist the elderly people to live both independently and actively should leverage on the efforts from both technical side and social side. We gladly notice that currently more and more efforts are being pursued on this direction, not only from the research community [3], but also from the government side: such efforts focus on the social connections between the assisted people and the outside world, such as witnessed e.g. by the second call of the European AAL programme, which stated the importance of helping the elderly people live actively and enjoy their life, bridging distances and preventing loneliness and isolation [14].

We believe human participation to be the key ingredient to being able to meet the challenges of building an efficient and effective AAL system. Indeed, human participation makes it possible to alleviate the sense of intrusion, to diversify the service categories, to explore the potential of the assistive devices, and also provides the elderly people with chances to keep on serving our society and living actively. Unfortunately, although researchers are realizing more and more the importance of human participation, due to implementation difficulties, this key ingredient – people, actually – appears to be not enough exploited in many of the current solutions.

Mutual assistance community, where people mutually assist each other, is an example of an architecture for AAL and a possible approach to bring human services into AAL systems. Smart devices can still be used in such a community to guarantee the safety of elderly people. Elderly people can also actively maintain their social networks, and regain their self-esteem. We are convinced that bringing the people – the missing ones! – in home-care environment is both efficacious and effective in saving the social resources, providing timely services and helping the elderly people live in an active and thus more satisfactory way.

This paper presented the structure of our proposed mutual assistance community as well as some of its basic components, e.g. service orchestration. Scenarios and simulation results were also provided, which demonstrated the effectiveness of human participation.

## Acknowledgement

The authors would like to thank IBM for awarding the IBM PhD Fellowship to Hong Sun to support his research in this domain.